\documentclass[12pt]{article}

\newcommand{\blind}{0}

\usepackage{tikz, adjustbox, float} 
\usepackage[round]{natbib} 
\usepackage{amsmath, amsfonts, amssymb, amsthm, dsfont} 
\usepackage{mathrsfs} 
\usepackage{mathabx} 
\usepackage{soul} 
\usepackage{bm} 
\usepackage{booktabs}  
\usepackage{enumitem} 
\usepackage{algorithm,algorithmic}

\def\real{{\rm I\!R}}

\def\0{{\bf 0}}
\def\1{{\bf 1}}

\def\bOmega{{\bm{\Omega}}}

\def\x{{\bf x}}
\def\I{{\bf I}}

\def\B{{\bf B}}

\DeclareMathOperator*{\var}{var}

\DeclareMathOperator*{\argzero}{argzero}

\usepackage{latexsym}
\usepackage{amsthm}
\usepackage{amsmath,amssymb,amsfonts,bm,amsbsy,bbm}
\usepackage{epsfig}
\usepackage{graphicx,rotating,lscape}
\usepackage{hyperref}
\usepackage{verbatim}
\usepackage{natbib}
\usepackage{multirow,color}
\usepackage{geometry}
\usepackage{setspace}
\usepackage{subfigure}
\usepackage{bbm}
\usepackage{ulem} 

\def\x{{\bf x}}
\def\U{{\bf U}}

\def\s{{\bf s}}
\def\a{{\bf a}}

\def\f{{\bf f}}

\def\bt{{\boldsymbol\theta}}
\def\bT{{\boldsymbol\Theta}}
\def\bb{{\boldsymbol\beta}}

\def\bx{\boldsymbol\xi}

\def\0{{\bf 0}}
\def\trans{^{\rm T}}
\def\pr{\hbox{Pr}}
\def\wh{\widehat}
\def\wt{\widetilde}
\def\wc{\widecheck}

\def\var{\hbox{var}}

\def\bse{\begin{eqnarray*}}
\def\ese{\end{eqnarray*}}
\def\be{\begin{eqnarray}}
\def\ee{\end{eqnarray}}
\def\bsq{\begin{equation*}}
\def\esq{\end{equation*}}
\def\bq{\begin{equation}}
\def\eq{\end{equation}}

\def\boxit#1{\vbox{\hrule\hbox{\vrule\kern6pt  \vbox{\kern6pt#1\kern6pt}\kern6pt\vrule}\hrule}}
\def\bse{\begin{eqnarray*}}
\def\ese{\end{eqnarray*}}
\def\be{\begin{eqnarray}}
\def\ee{\end{eqnarray}}
\def\bsq{\begin{equation*}}
\def\esq{\end{equation*}}
\def\bq{\begin{equation}}
\def\eq{\end{equation}}
\def\th{^{th}}
\def\var{\hbox{var}}

\def\wh{\widehat}
\def\wt{\widetilde}

\def\th{^{\rm th}}

\def\trans{^{\rm T}}

\def\bg{{\boldsymbol\gamma}}
\def\bG{{\boldsymbol\Gamma}}

\def\a{{\bf a}}
\def\B{{\bf B}}

\def\d{{\bf d}}

\def\f{{\bf f}}

\def\I{{\bf I}}

\def\U{{\bf U}}

\def\v{{\bf v}}

\def\w{{\bf w}}
\def\X{{\bf X}}

\def\x{{\bf x}}

\def\bSig{{\bf \Sigma}}

\def\log{\hbox{log}}

\def\squarebox#1{\hbox to #1{\hfill\vbox to #1{\vfill}}}

\def\balpha{{\boldsymbol \alpha}}

\def\0{{\bf 0}}

\def\var{\hbox{var}}

\def\pr{\hbox{Pr}}
\def\wh{\widehat}
\def\wt{\widetilde}

\def\log{\hbox{log}}

\newtheoremstyle{mytheoremstyle} 
    {0.3cm}                      
    {0cm}                        
    {}
    {}                           
    {\bf}                   
    {: }                          
    {0em}                       
    {}  

\theoremstyle{mytheoremstyle}
\newtheorem{Theorem}{Theorem}

\newtheorem*{Lemma*}{Lemma}
\newtheorem{Corollary}{Corollary}
\newtheorem{Proposition}{Proposition}
\newtheorem{Assumption}{Assumption}

\newtheoremstyle{myExampleRemarkstyle} 
    {0.3cm}                    
    {0cm}                      
    {\it}                         
    {}                         
    {\bf}                      
    {: }                       
    {0em}                      
    {}  

\theoremstyle{myExampleRemarkstyle}

\newtheorem{innercustomgeneric}{\customgenericname}
\providecommand{\customgenericname}{}
\newcommand{\newcustomtheorem}[2]{%
  \newenvironment{#1}[1]
  {%
   \renewcommand\customgenericname{#2}%
   \renewcommand\theinnercustomgeneric{##1}%
   \innercustomgeneric
  }
  {\endinnercustomgeneric}
}

\newcustomtheorem{customthm}{Theorem}
\newcustomtheorem{customlemma}{Lemma}
\usepackage{mathalfa}
\usepackage{oubraces}

\let\refBKP\ref
\renewcommand{\ref}[1]{{\upshape\refBKP{#1}}}

\usepackage{setspace}
\usepackage{xr}
\makeatletter
\newcommand*{\addFileDependency}[1]{
  \typeout{(#1)}
  \@addtofilelist{#1}
  \IfFileExists{#1}{}{\typeout{No file #1.}}
}
\makeatother

\begin{document}

\begingroup  
\begin{center}
\LARGE 
Calibrated Estimation and Inference for \\ Semiparametric Regression Models
\if1\blind\else
\vspace{0.2cm}\\
\normalsize{Yuming Zhang$^{1}$, Yanyuan Ma$^{2}$, Tianxi Cai$^{1}$, Xuming He$^{3}$, St\'ephane Guerrier$^{4,5}$}
\vspace{0.2cm}\\
\footnotesize{$^{1}$Department of Biostatistics, Harvard University, USA \\ $^{2}$Department of Statistics, Pennsylvania State University, USA \\ $^{3}$Department of Statistics and Data Science, Washington University in St. Louis, USA \\ $^{4}$School of Pharmaceutical Sciences, Faculty of Science, University of Geneva, Switzerland \\ $^{5}$Section of Earth and Environmental Sciences, Faculty of Science, University of Geneva, Switzerland}
\fi
\end{center}
\endgroup

\vspace{-0.65cm}  
\begin{abstract}
\vspace{-0.25cm} 

We consider a broad class of semiparametric regression models in which the conditional distribution of the response takes the form $f\{Y|\x\trans\bb+m(z),\phi\}$, known up to a parametric component $\bb$ of diverging dimension $p$, a smooth function $m(\cdot)$, and a dispersion parameter $\phi$. The existing literature on such models has focused on semiparametric efficiency for $\bb$, treating $\phi$ and $m(\cdot)$ as nuisances and largely ignoring finite-sample bias. Yet this bias can be substantial, particularly when $p$ is large relative to $n$ or the dispersion is high, and it can seriously undermine inference for $\bb$; moreover, $\phi$ is often of direct scientific interest. We therefore propose SABRE, a general calibration framework for semiparametric estimation and inference, which calibrates an initial estimator against its model-implied expectation under a tractable parametric approximation to the semiparametric model. For generalized partially linear models, we show that SABRE reduces the bias of both $\bb$ and $\phi$, accommodates a diverging parameter dimension without sparsity, and preserves the first-order variance and semiparametric efficiency of the initial estimator; the joint construction also improves estimation and inference for $m(\cdot)$. Simulation studies and an application to Alzheimer's disease genetics association analysis demonstrate the empirical effectiveness of SABRE in reducing bias and improving inference.

\end{abstract}

\newpage

\section{Introduction}
\label{sec:intro}

\subsection{Background}

Statistical analyses in real-world applications often require flexible regression models that can capture complex relationships between responses and covariates, while retaining a finite-dimensional parameter that yields a clear interpretation of the covariate-response association. Motivated by this, we consider a broad class of semiparametric models in which the conditional distribution of $Y|\x,z$ takes the form $f\{Y|\x\trans\bb+m(z),\phi\}$, with $\bb$, $m(\cdot)$, and $\phi$ all unknown, yielding mean $\mathbb{E}(Y|\x,z) = g\{\x\trans\bb+m(z)\}$ and variance $\var(Y|\x,z) = v\{\x\trans\bb+m(z),\phi\}$ for known functions $g(\cdot)$ and $v(\cdot,\cdot)$. The covariates $(\x,z)$ are deterministic, with $\x\in\real^p$ and $z\in[0,1]$. The parametric component $\bb$ has dimension $p$, which is allowed to grow with the sample size $n$. The nonparametric component $m(\cdot):[0,1]\to\mathcal{M}$ is a smooth function taking values in a bounded set $\mathcal{M}\subset\real$; for identifiability, the intercept is absorbed into $m(\cdot)$. Finally, $\phi\in\Phi\subset(0,\infty)$ is a dispersion parameter. This model class includes the generalized partially linear models (GPLMs; see e.g., \citealt{boente2006robust}). 

Our model class encompasses many settings that arise frequently in practice. In studies based on electronic health records (EHRs), the covariates comprise a large number of diagnosis, procedure, and medication codes; in label-only differential privacy (label-DP), they encode numerous user- and context-specific features. In both, $p/n$ is typically not small, making a diverging parameter dimension the realistic asymptotic regime. Moreover, the scientific target is a latent response $Y^\circ$ (e.g., a true disease status, or a true label prior to privatization) that follows a regression with linear predictor $\x\trans\bb+m(z)$ but is never observed. What one observes instead is a corrupted version $Y$, such as an EHR-derived phenotype (e.g., \citealt{lu2024leveraging}) or a thresholded biomarker (e.g., \citealt{cai2004semi}), a randomized label (e.g., \citealt{warner1965randomized,ghazi2021deep}), a misclassified disease status in epidemiology (e.g., \citealt{chen2013effects}), a noisy case-control label in genetic analyses (e.g., \citealt{biffani2017effect}), or a self-reported response in public health surveys (e.g., \citealt{preston2015effects}). Because the corruption mechanism is treated as known, either from domain knowledge or estimated from external data and then regarded as fixed, the observed $Y$ again follows $f\{Y|\x\trans\bb+m(z),\phi\}$, determined jointly by the latent model for $Y^\circ$ and this mechanism, and hence known up to $\bb$, $m(\cdot)$, and $\phi$.

Across these settings, $\bb$ is the primary target for estimation and inference, characterizing the association between the covariates and the latent $Y^\circ$ rather than the observed, noisy $Y$. The dispersion $\phi$ is also often of direct scientific interest and plays a key role in studies across biomedical research (e.g., \citealt{breslow1984extra}), genomics (e.g., \citealt{anders2010differential}), insurance and actuarial science (e.g., \citealt{smyth2002fitting}), as well as in sample size calculations based on preliminary analyses such as pilot studies.

The semiparametric literature on related models has focused on estimators for $\bb$ that attain semiparametric efficiency, while treating $\phi$ and $m(\cdot)$ as nuisances (e.g., \citealt{murphy2000profile,hardle2004nonparametric,jankova2018semiparametric}). Their bias is typically overlooked as asymptotically negligible relative to the standard error, yet in finite samples it can be large enough to distort confidence intervals (CIs) and hypothesis tests, as is well documented for maximum likelihood estimators (MLEs) in parametric models (e.g., \citealt{wang1999bias,mittelhammer2005empirical,sur2019modern}). Bias in $\phi$ compounds the problem, undermining inference on $\bb$ even when $\bb$ itself is estimated without bias (e.g., \citealt{kosmidis2014bias,heller2019beyond}). Bias correction for models of the form $f\{Y|\x\trans\bb+m(z),\phi\}$ is nonetheless absent: existing methods either correct bias in low-dimensional parametric models with fixed $p$, or pursue semiparametric efficiency without addressing bias.

In this paper, we close this gap with \textit{SemipArametric Bias-Reduced Estimation} (SABRE), a unified bias correction framework for this model class that simultaneously (i) yields bias-reduced estimators of $\bb$ and $\phi$, (ii) leaves the variability of $\bb$ unchanged, and (iii) remains valid when $p^2\log(n)n^{-1}\to 0$, thereby delivering more accurate and reliable finite-sample inference across the applications above.

\subsection{Related Works}

The model class considered here is well established and widely used in practice because it accommodates flexible nonlinear effects while retaining an interpretable linear component (e.g., \citealt{heckman1986spline,severini1994quasi}). However, the existing literature targets $\sqrt{n}$-consistency and semiparametric efficiency for low-dimensional $\bb$ rather than bias. Bias correction has a rich literature of its own, but almost entirely for parametric models. One line is \textit{preventive}, analytically modifying score equations to remove leading-order bias (e.g., \citealt{Firt:93,KoFi:09}); these adjustments are model-specific, must be re-derived for each likelihood, and are therefore difficult to generalize across a broad model class. A second line is \textit{corrective}, starting from an initial estimator and calibrating it via an estimated bias function, often through simulation-based schemes such as the bootstrap \citep{efron1994introduction}, classical minimum distance \citep{newey1994large}, or indirect inference \citep{gourieroux1993indirect}, whose bias properties have been studied in various parametric settings (e.g., \citealt{mackinnon1998approximate,gourieroux200013,guerrier2019simulation}). Because these corrective methods approximate expectations of the initial estimator by Monte Carlo averages under the exact parametric model, they are fundamentally parametric tools. Bias correction for the dispersion parameter is even scarcer, and confined to simple parametric models (e.g., \citealt{saha2005bias} for two-parameter negative binomial models).

A separate literature debiases penalized estimators such as the Lasso \citep{tibshirani1996regression} to enable valid inference for low-dimensional components of $\bb$ when $p$ can exceed $n$ (e.g., \citealt{zhang2014confidence,van2014asymptotically,ning2017general}). These methods are formulated for parametric generalized linear models (GLMs) and accommodate neither a functional component $m(\cdot)$, nor bias in $\phi$, nor a response that is a noisy manifestation of a latent truth.

\subsection{Contributions}

SABRE is a general estimator-calibration framework: it begins with an estimator obtained from a tractable parametric approximation to the semiparametric model and calibrates it by matching the observed estimate to its model-implied expectation. Because that expectation is rarely available in closed form, we evaluate it by simulation under the fitted model. Simulation provides a practical means for evaluation of the expectation, rather than defining the scope of the methodology.

To our knowledge, SABRE is the first bias correction method for the considered semiparametric model class. The parametric approximation that makes the expectation accessible despite the unknown $m(\cdot)$ is also what makes the theory very challenging. The approximation error must be carried through the analysis rather than assumed away; undersmoothed B-splines keep it from propagating into the parametric component, yet controlling it requires structurally different arguments for continuous and discrete responses. Moreover, $\bb$ and the spline coefficients representing $m(\cdot)$ converge at different rates and must be handled separately. Bounding the residual error requires a convergence rate for the simulated estimator that holds uniformly in the parameter, and a diverging $p$ makes this uniformity much harder to obtain. Compounding these, $\bb$ and $\phi$ solve a coupled system in which each calibration equation is indexed by the other, making their estimation errors intrinsically dependent. We develop the full theory for GPLMs and provide a roadmap for the broader class, where the methodology also applies and our numerical experiments show promising performance.

SABRE corrects bias in both $\bb$ and $\phi$ while allowing $p$ to diverge at a rate of $p^2\log(n)n^{-1}\to 0$, without sparsity assumptions or penalized methods. We also obtain a bias-reduced estimator of $m(\cdot)$; when $m(\cdot)$ is itself the target, we introduce an \textit{optimal SABRE} variant which attains standard properties such as minimum mean squared error (MSE) without requiring undersmoothing. The correction reduces the bias order while retaining the same asymptotic variance as the initial estimator, so bias reduction is achieved without sacrificing semiparametric efficiency for $\bb$. Finally, because SABRE requires only the ability to generate responses under the fitted model, rather than model-specific analytical bias formulas, it accommodates complex observed-response mechanisms (e.g., misclassified and privatized responses) for which analytical corrections are unavailable or difficult to derive.

Simulation studies and a real-world genetic association study of Alzheimer's disease (AD) in an EHR-linked biobank demonstrate substantial finite-sample gains with SABRE, which markedly reduces bias in $\bb$, $\phi$, and $m(\cdot)$ relative to alternative methods. The improvements are most pronounced when $p/n$ is large and/or dispersion is high, and translate into CIs for $\bb$ with more accurate empirical coverage and typically shorter length. In the AD study, where an error-prone diagnostic label surrogates for true disease status and a large panel of genetic variants makes $p/n$ large, SABRE recovers well-established genetic risk factors and corrects the inflated effect sizes that standard methods produce.

\subsection{Organization and Notation}

Section~\ref{sec:methodology} introduces the B-spline approximating model, the initial estimator, and the SABRE methodology. Section~\ref{sec:theory} gives the roadmap for the asymptotic theory and develops it within the GPLM subclass. Sections~\ref{sec:simulations} and~\ref{sec:case_study} report respectively the simulation studies and the AD genetics application, and Section~\ref{sec:conclusion} concludes. Proofs and additional materials are collected in the supplementary material.

We define some notation used throughout. Let $[n]\equiv \{1,\ldots,n\}$. For positive real numbers $a_n$ and $b_n$, we write $a_n=o(b_n)$ or $a_n\ll b_n$ if $\lim_{n\to\infty} a_n/b_n = 0$, and write $a_n=\mathcal{O}(b_n)$ or $a_n \lesssim b_n$ if $a_n\leq C b_n$ for a large enough $n$ and some finite positive $C$. We write $a_n\asymp b_n$ if $a_n = \mathcal{O}(b_n)$ and $b_n = \mathcal{O}(a_n)$. For a vector $\a = (a_1,\ldots, a_d)\trans \in \real^d$, let $\|\a\|_q \equiv (|a_1|^q + \ldots + |a_d|^q)^{1/q}$ with $q\in\mathbb{N}^+$ and $\|\a\|_\infty \equiv \max_{i\in[d]}|a_i|$. We denote the set of the $q\th$ order smooth functions as $C^q([0,1])\equiv \{g: g^{(q)} \in C([0,1])\}$, where $C([0,1])$ denotes the set of continuous functions on $[0,1]$. $\lambda_{\min}(\cdot)$ and $\lambda_{\max}(\cdot)$ denote minimum and maximum eigenvalues, respectively. Finally, we define $\argzero_{\bt\in\bT} \f(\bt) \equiv \{\bt\in\bT: \f(\bt) = \0\}$.

\section{Methodology}
\label{sec:methodology}

\subsection{B-spline Approximating Model and Initial Estimator}

To start, we approximate the unknown function $m(\cdot)$ with B-splines: 
\bq \label{eqn:Bsplines_approx}
    m(z) \approx \B(z)\trans \balpha = \sum_{j=1}^K \alpha_j B_j(z)
    \quad \text{with} \quad z\in[0,1], 
\eq
where $\{B_j(z)\}_{j\in[K]}$ are B-spline basis functions of order $r$, $K$ is the number of basis functions, and $\balpha=(\alpha_1,\ldots,\alpha_K)\trans$ is the vector of spline coefficients. Let $N$ denote the number of interior knots, and define the knot sequence 
\bsq
    t_1=\ldots=t_r= 0< t_{1+r}<\ldots<t_{N+r}<1= t_{N+r+1}=\ldots=t_{N+2r}.
\esq
With the intercept absorbed into $m(\cdot)$, we have $K= N+r$ and $ \sum_{j=1}^K B_j(z)=1$ for all $z\in [0,1]$. B-splines are widely used for nonparametric and semiparametric estimation due to their numerical efficiency and well-understood asymptotic properties (see e.g., \citealt{de1978practical,huang2003local,wang2009spline,stone1985additive,huang2002varying,huang2004identification,liu2011estimation,wang2011estimation,ma2017semiparametric}). Under suitable conditions specified in Section~\ref{sec:theory}, there exists some $\balpha$ such that $\B(\cdot)\trans\balpha$ is sufficiently close to $m(\cdot)$.

We collect the regression and spline coefficients in $\bg \equiv (\bb\trans, \balpha\trans)\trans\in\bG\subset\real^{p+K}$, and the covariates and B-spline basis functions in $\w_i\equiv \{\x_i\trans, \B(z_i)\trans\}\trans$. Let $\nu_i^*(\bg)\equiv \w_i\trans\bg$ denote the linear predictor for the $i\th$ observation under the B-spline approximation. The true model $f\{Y_i| \x_i\trans\bb+m(z_i), \phi\}$ can then be approximated by $f\{Y_i| \nu_i^*(\bg), \phi\}$, with mean $\mu_i^*(\bg) \equiv g\{\nu_i^*(\bg)\}$ and variance $\var^*(Y_i|\x_i,z_i) \equiv v\{\nu_i^*(\bg), \phi\}$. 

We define the score for $\bg$ as 
\bsq
    \U_\bg\left(\bg,\phi; \{Y_i\}_{i\in[n]}\right) \equiv \sum_{i=1}^n \frac{\partial \log f\{Y_i|\nu_i^*(\bg), \phi\}}{\partial \nu_i^*(\bg)} \w_i.
\esq
As an estimating equation for $\phi$, we use a Pearson-type moment condition based on squared residuals: 
\bsq
    \U_\phi\left(\bg,\phi; \{Y_i\}_{i\in[n]}\right) \equiv \sum_{i=1}^n \left[\frac{\{Y_i - \mu_i^*(\bg)\}^2}{v\{\nu_i^*(\bg), \phi\}} - 1\right].
\esq
We then define the \textit{B-spline MLE} (sMLE), $\wt{\bg} = (\wt{\bb}\trans, \wt{\balpha}\trans)\trans$ and $\wt{\phi}$, as the joint solution of these equations 
\bq \label{eqn:def:gamma_phi_tilde}
    \begin{bmatrix}
        \wt{\bg} \\ \wt{\phi}
    \end{bmatrix} \equiv \underset{\bg\in\bG, \phi\in\Phi}{\argzero}\; \begin{bmatrix}
        \U_\bg\left(\bg,\phi; \{Y_i\}_{i\in[n]}\right) \\ \U_\phi\left(\bg,\phi; \{Y_i\}_{i\in[n]}\right)
    \end{bmatrix},
\eq
and the corresponding sMLE of $m(\cdot)$ is $\wt{m}(\cdot) \equiv
\B(\cdot)\trans\wt{\balpha}$.

The sMLE $\{\wt{\bb}, \wt{\phi},\wt{m}(\cdot)\}$ is a natural choice of initial estimator for SABRE. Indeed, under suitable conditions stated later, $\wt{\bb}$ is consistent, asymptotically normal, and semiparametrically efficient, while $\wt{\phi}$ is consistent. These properties make sMLE a suitable starting point for our bias correction procedure developed in the next section.

\subsection{The SABRE Methodology}

As discussed in Section~\ref{sec:intro}, SABRE is a simulation‑based framework that relies on simulating data from a parametric approximating model that is ``close'' to the true model. We therefore distinguish carefully between (i) the observed data from the true model, (ii) synthetic data from the B‑spline approximating model, and (iii) the estimators computed from each. Throughout, we write $\bt \equiv \{\bb\trans, m(\cdot)\}\trans$, and use $\bb_0,\phi_0,m_0(\cdot)$ to denote the true parameters. The observed response at covariates $(\x_i,z_i)$ from the true model with $(\bt_0, \phi_0)$ is $Y_i$, and we use $Y_i^*(\bg,\phi)$ for a synthetic response generated at covariates $\w_i=\{\x_i\trans, \B(z_i)\trans\}\trans$ from the B-spline approximating model with generic $(\bg,\phi)$. Expectations under the true model are written as $\mathbb{E}(\cdot)$, whereas expectations under the B‑spline approximating model with $(\bg,\phi)$ are written as $\mathbb{E}_{\bg,\phi}(\cdot)$.

The sMLE $(\wt{\bg}\trans,\wt{\phi})\trans$ in
\eqref{eqn:def:gamma_phi_tilde} are computed from the observed $\{Y_i\}_{i\in[n]}$. For any given approximating-model parameter $(\bg,\phi)$, we define the parametric counterpart as 
\bq \label{eqn:def:gamma_phi_tilde_star}
    \begin{bmatrix}
        \wt{\bg}^*(\bg,\phi) \\ \wt{\phi}^*(\bg,\phi)
    \end{bmatrix} \equiv \underset{\bx\in\bG, \delta\in\Phi}{\argzero}\; \begin{bmatrix}
        \U_\bg\left(\bx,\delta; \{Y_i^*(\bg,\phi)\}_{i\in[n]}\right) \\ \U_\phi\left(\bx,\delta; \{Y_i^*(\bg,\phi)\}_{i\in[n]}\right)
    \end{bmatrix},
\eq
that is, the sMLE applied to the simulated responses $\{Y_i^*(\bg,\phi)\}_{i\in[n]}$. We then define the SABRE estimator, $\wh{\bg} = (\wh{\bb}\trans, \wh{\balpha}\trans)\trans$ and $\wh{\phi}$, as the joint solution of these equations:
\bq \label{eqn:def:gamma_phi_hat}
    \begin{bmatrix}
        \wh{\bg} \\ \wh{\phi}
    \end{bmatrix} \equiv \underset{\bg\in\bG, \phi\in\Phi}{\argzero}\; \begin{bmatrix}
        \wt{\bg} - \mathbb{E}_{\bg,\phi}\left\{\wt{\bg}^*(\bg,\phi)\right\} \\ \wt{\phi} - \mathbb{E}_{\bg,\phi}\left\{\wt{\phi}^*(\bg,\phi)\right\}
    \end{bmatrix},
\eq
and the corresponding SABRE estimator of $m(\cdot)$ is $\wh{m}(\cdot) \equiv \B(\cdot)\trans\wh{\balpha}$. 

Intuitively, $(\wt{\bg}\trans, \wt{\phi})\trans$ is a functional of the observed data that ``reflects'' the true data‑generating process, whereas $\{\wt{\bg}^*(\bg,\phi)\trans, \wt{\phi}^*(\bg,\phi)\}\trans$ is the same functional computed on data generated from the B‑spline approximating model with parameter $(\bg,\phi)$. If the approximating model is well-specified and ``close'' to the true semiparametric model, then the value of $(\bg\trans,\phi)\trans$ for which $(\wt{\bg}\trans,\wt{\phi})\trans$ and $\mathbb{E}_{\bg, \phi}[\{\wt{\bg}^*(\bg,\phi)\trans, \wt{\phi}^*(\bg,\phi)\}\trans]$ agree is a good candidate for the true parameter, leading to SABRE. This is analogous to \textit{method of moments}, where one matches sample moments (functionals of the observed data) to their model‑implied counterparts, and to \textit{simulated method of moments} \citep{mcfadden1989method}, where the model‑implied moments are approximated by simulation when they are not analytically tractable. It also relates to \textit{indirect inference} \citep{gourieroux1993indirect} in parametric models, which chooses parameters so that an auxiliary estimator computed on the observed data agrees with the same auxiliary estimator computed on simulated data.

In practice, the expectations in \eqref{eqn:def:gamma_phi_hat} have no closed form. Because they are taken under the parametric B‑spline approximating model, they can be approximated by averaging sMLEs computed on data simulated from that model at a given $(\bg,\phi)$. Solving \eqref{eqn:def:gamma_phi_hat} by generic optimization, however, would require re‑evaluating these Monte Carlo expectations at every step, which is computationally burdensome when the model dimension is large. We instead employ \textit{stochastic approximation methods} (e.g., \citealt{kushner2003stochastic}), which solve \eqref{eqn:def:gamma_phi_hat} from simulated samples without explicitly forming the expectations. Empirically, convergence is fast and stable even when $p$ and $n$ are both relatively large. Further implementation details are provided in Supplement~A.


\section{Theory}
\label{sec:theory}

\subsection{B-spline Approximation}

SABRE heavily relies on the fact that the B-spline approximating model is ``close'' to the true semiparametric model. In this section, we lay out conditions under which this holds and quantify what we mean by ``close''. 

\setcounter{Assumption}{0}
\renewcommand\theAssumption{S.\arabic{Assumption}}

\begin{Assumption}
\label{assum:spline:1}
The true function $m_0(\cdot)$ satisfies $m_0(\cdot) \in C^q([0,1])$ with $q \geq 2$, and the spline order $r$ satisfies $r \geq q$.
\end{Assumption}

\begin{Assumption}
\label{assum:spline:3}
The number of interior knots $N$ diverges with $n$. Moreover, it satisfies $N^2 \{\log(n)\}^2n^{-1} \to 0$ and $n^{3/2}N^{-2q} = \mathcal{O}(1)$. 
\end{Assumption}

\begin{Assumption}
\label{assum:spline:4}
Let $h_j$ be the distance between neighboring knots as $h_j\equiv t_{j+1}-t_{j}$ with $r\leq j\leq N+r$. We also let $h\equiv\max_{r\leq j \leq N+r} h_j$ and $h'\equiv\min_{r\leq j \leq N+r} h_j$. There exists some finite positive constant $C_h$ such that $h/h'\leq C_h$. 
\end{Assumption}

Assumptions~\ref{assum:spline:1} and \ref{assum:spline:4} are standard conditions in the B-splines literature (e.g., \citealt{shen1998local,ma2017semiparametric,jiang2018spline,lee2023semiparametric}). Assumption~\ref{assum:spline:3} requires undersmoothed B-splines, which is frequently imposed to control the bias of the nonparametric component estimation in order to facilitate the estimation properties of the parametric component of interest. Under Assumptions~\ref{assum:spline:1} to \ref{assum:spline:4}, there exists some $\balpha_0$ such that $\B(\cdot)\trans\balpha_0$ is sufficiently close to $m_0(\cdot)$ and 
\bq \label{eqn:order_true_function_diff}
    \sup_{z\in[0,1]} \left|\B(z)\trans\balpha_0 - m_0(z)\right| = \mathcal{O}(h^q).
\eq
This is a standard result in the literature \citep{de1978practical} and is commonly used in the nonparametric smoothing literature (e.g., \citealt{shen1998local,liu2011estimation,ma2017semiparametric,lee2023semiparametric}). This result suggests that the B-spline approximation error is negligible for sufficiently smooth functions, which can be achieved through putting sufficiently many knots that are more or less evenly spread.

\subsection{Roadmap and Scope of the Asymptotic Theory for SABRE}

The asymptotic properties we aim to establish for SABRE are: (i) convergence rates of $\wh{\bg}$ and $\wh{\phi}$, (ii) asymptotic normality of $\wh{\bg}$, and (iii) bias orders of $\wh{\bg}$ and $\wh{\phi}$. Properties of $\wh{\bg}$ immediately yield corresponding results for $\wh{\bb}$ and $\wh{m}(\cdot)$. The theoretical investigation proceeds in three steps:
\begin{itemize}
    \item Step~1: Fix a preliminary estimator $\wc{\phi}$ for $\phi_0$, and let
    \bq \label{eqn:def:gamma_hat}
        \wh{\bg}(\wc{\phi}) \equiv \argzero_{\bg\in\bG}\; \left[ \wt{\bg} - \mathbb{E}_{\bg, \wc{\phi}} \{\wt{\bg}^*(\bg,\wc{\phi})\} \right].
    \eq 
    Under mild conditions on $\wc{\phi}$, we establish the asymptotic properties of $\wh{\bg}(\wc{\phi})$. 
    
    \item Step~2: Treating $\wh{\bg}(\wc{\phi})$ as fixed, we define 
    \bq \label{eqn:def:phi_hat}
        \wh{\phi}(\wc{\phi}) \equiv \argzero_{\phi\in\Phi}\; \left( \wt{\phi} - \mathbb{E}_{\wh{\bg}(\wc{\phi}), \phi}\left[\wt{\phi}^*\{\wh{\bg}(\wc{\phi}), \phi\}\right] \right),
    \eq
    and establish its asymptotic properties.  
    
    \item Step~3: We show that $\wh{\phi}(\wc{\phi})$ from Step~2 satisfies the same conditions imposed on $\wc{\phi}$ in Step~1. Consequently, since $\wh{\bg}$ and $\wh{\phi}$ are taken as the joint solution to~\eqref{eqn:def:gamma_phi_hat}, they inherit the asymptotic properties of $\wh{\bg}(\wc{\phi})$ and $\wh{\phi}(\wc{\phi})$ in Steps~1 and~2, respectively.
\end{itemize}
The key step is therefore to establish the asymptotic properties of $\wh{\bg}(\wc{\phi})$ and $\wh{\phi}(\wc{\phi})$ under suitable conditions on $\wc{\phi}$. We outline the proof strategy and the main technical challenges for $\wh{\bg}(\wc{\phi})$, abbreviated as $\wh{\bg}^\circ$; the analysis of $\wh{\phi}(\wc{\phi})$ is analogous.

Let $\wt{\bg}^* \equiv \wt{\bg}^*(\bg_0,\phi_0)$ denote the sMLE computed on synthetic data $\{Y_i^*(\bg_0,\phi_0)\}_{i\in[n]}$ from the B-spline approximating model at the parameter value closest to the truth; it is introduced only for analysis and is not accessible in practice. Using $\wt{\bg} = \mathbb{E}_{\wh{\bg}^\circ, \wc{\phi}}\{\wt{\bg}^*(\wh{\bg}^\circ, \wc{\phi})\}$ from \eqref{eqn:def:gamma_hat}, we have
\bq \label{eqn:decomposition_gamma_hat}
    \wh{\bg}^\circ - \bg_0 = 
    \underbrace{\wt{\bg} - \wt{\bg}^*}_{\text{part~1}} + \underbrace{\wt{\bg}^*-\bg_0}_{\text{part~2}} + \underbrace{\wh{\bg}^\circ - \mathbb{E}_{\wh{\bg}^\circ, \wc{\phi}}\left\{\wt{\bg}^*(\wh{\bg}^\circ, \wc{\phi})\right\}}_{\text{part~3}}.
\eq

Part~1 captures the discrepancy induced by replacing the true semiparametric model with a B‑spline approximating model. It is negligible when the B‑spline approximation error is small, but controlling it formally is demanding: the argument depends on whether the response is continuous or discrete, and the two blocks $\bb$ and $\balpha$ of $\bg$ converge at different rates and must be handled separately. Part~2 summarizes the behavior of the sMLE under the B-spline approximating model at $(\bg_0,\phi_0)$, and is the most tractable term since it involves only a parametric model. Part~3 measures the residual error of $\wt{\bg}^*(\wh{\bg}^\circ,\wc{\phi})$ under the approximating model at \((\wh{\bg}^\circ,\wc{\phi})\); it is also negligible, yet bounding it requires the convergence rate of $\wt{\bg}^*(\bg,\wc{\phi})-\bg$ uniformly in $\bg$, which is non-trivial under a diverging model dimension. The key implication is that, once parts~1 and~3 are shown to be of smaller order, $\wh{\bg}^\circ$ inherits the convergence rate and asymptotic distribution of $\wt{\bg}^*$, and hence the same efficiency for $\bb$, as we establish later.

To study the bias of $\wh{\bg}^\circ$, we decompose
\bsq
    \mathbb{E}\left(\wh{\bg}^\circ\right) - \bg_0 = \underbrace{\mathbb{E}\left(\wt{\bg}-\wt{\bg}^*\right)}_{\text{part~1}} + \underbrace{\mathbb{E}\left\{\v^*(\bg_0,\phi_0)\right\}}_{\text{part~2}} + \underbrace{\d^*(\bg_0,\phi_0) - \mathbb{E}\left\{\d^*(\wh{\bg}^\circ,\wc{\phi})\right\}}_{\text{part~3}},
\esq
where, given generic parameter $(\bg,\phi)$, we define $\d^*(\bg,\phi) \equiv \mathbb{E}_{\bg,\phi}\{\wt{\bg}^*(\bg,\phi)\} - \bg$ as the finite-sample bias of $\wt{\bg}^*(\bg,\phi)$, and $\v^*(\bg_0,\phi_0)\equiv \wt{\bg}^* - \mathbb{E}_{\bg_0,\phi_0}(\wt{\bg}^*)$ as the error vector. Parts~1 and~2 again reflect the approximating-vs-true model discrepancy and are of smaller order, so part~3 dominates. This is why $\wh{\bg}^\circ$ is less biased than the sMLE $\wt{\bg}$: the bias of $\wt{\bg}$ is driven by $\d^*(\bg_0,\phi_0)$ alone, whereas that of $\wh{\bg}^\circ$ is driven by the difference $\d^*(\bg_0,\phi_0) - \mathbb{E}\{\d^*(\wh{\bg}^\circ,\wc{\phi})\}$, which a Taylor expansion shows to be of strictly smaller order under smoothness of $\d^*(\bg,\phi)$ and convergence of $(\wh{\bg}^\circ,\wc{\phi})$ to $(\bg_0,\phi_0)$. In effect, the SABRE correction cancels the leading term in the bias of $\wt{\bg}$, leaving only higher-order remainders.

The proof strategies outlined above apply to the broad semiparametric model class $f\{Y|\x\trans\bb+m(z),\phi\}$, to which SABRE is applicable. However, given the technical complexity and to keep the presentation focused, in the rest of this section we develop the asymptotic theory within the subclass of GPLMs, where $f$ takes the exponential family form:
\bq \label{eqn:def_GPLM}
    f\{y|\x\trans\bb + m(z),\phi\} = \exp\left\{\frac{y\nu - b(\nu)}{\phi}  + c(y,\phi)\right\} \quad \text{with} \quad \nu = \x\trans\bb + m(z),
\eq
and $b(\cdot),c(\cdot,\cdot)$ are known. The mean and variance are $\mathbb{E}(Y|\x,z)=b'(\nu)$ and $\var(Y|\x,z)=\phi b''(\nu)$ respectively, where $b'(\cdot),b''(\cdot)$ denote the first and second derivatives of $b(\cdot)$.

\subsection{Asymptotic Properties}

We state the rest of the assumptions needed for the asymptotic theory.

\renewcommand{\theAssumption}{P.\arabic{Assumption}}\setcounter{Assumption}{0}

\begin{Assumption}
\label{assum:p}
The parameter dimension $p$ satisfies $p^2\log(n)n^{-1}\to 0$ as $n\to\infty$. 
\end{Assumption}

\begin{Assumption}
\label{assum:para_space}
The parameter space $\bG$ of $\bg$ is a compact and convex subset of
$\real^{p+K}$, and $\bg_0$ lies in the interior of $\bG$. Moreover, the parameter space $\Phi$ of $\phi$ is bounded and bounded away from zero.  
\end{Assumption}

Assumption~\ref{assum:p} allows $p$ to diverge with $n$ at a mild rate, as is common in parametric settings (e.g., \citealt{he2000parameters}). 
Assumption~\ref{assum:para_space} is a regularity condition to ensure that expansions can be made between $\bg_0$ and an arbitrary point in $\bG$. The condition on $\Phi$ guarantees the variance of the response to be bounded and bounded away from zero.

\renewcommand{\theAssumption}{C.\arabic{Assumption}} \setcounter{Assumption}{0}

\begin{Assumption}
\label{assum:covariate:x}
The covariates $\{\x_i\}_{i\in[n]}$ satisfy the followings:
\begin{enumerate}
    \item For sufficiently large $n$, $c_x \leq \lambda_{\min}(n^{-1}\sum_{i=1}^n \x_i\x_i\trans) \leq \lambda_{\max}(n^{-1}\sum_{i=1}^n \x_i\x_i\trans) \leq C_x$, with finite positive constants $c_x,C_x$. 
    \item $\sup_{\s\in\real^p: \|\s\|_2=1} \; \sum_{i=1}^n |\s\trans\x_i|^4 = \mathcal{O}(n)$. 
\end{enumerate}
\end{Assumption}

Assumption~\ref{assum:covariate:x} can be satisfied, for example, when $\{\x_i\}_{i\in[n]}$ can be viewed as realizations from a $p$-variate distribution where $\mathbb{E}\{(\s\trans\x_i)^2\}$ and $\mathbb{E}\{(\s\trans\x_i)^4\}$ are uniformly bounded for $\s\in\real^p$ such that $\|\s\|_2=1$.  

\begin{Assumption}
\label{assum:covariate:z}
The covariates $\{z_i\}_{i\in[n]}$ satisfy the followings:
\begin{enumerate}
    \item $\displaystyle\sup_{\s\in\real^K:\|\s\|_2=1}n^{-1}\sum_{i=1}^n\{\s\trans\B(z_i)\}^2 \asymp \sup_{\s\in\real^K:\|\s\|_2=1}
      \int_0^1\{\s\trans\B(z)\}^2 dz$. 
    \item 
      $\displaystyle\inf_{\s\in\real^K:\|\s\|_2=1}n^{-1}\sum_{i=1}^n\{\s\trans\B(z_i)\}^2
      \asymp \inf_{\s\in\real^K:\|\s\|_2=1}
      \int_0^1\{\s\trans\B(z)\}^2 dz$. 
\end{enumerate}
\end{Assumption}

Assumption~\ref{assum:covariate:z} can be satisfied, for example, when $\{z_i\}_{i\in[n]}$ can be viewed as realizations of iid random variables $\{Z_i\}_{i\in[n]}$ according to a probability density function distributed on $[0,1]$ that is bounded and bounded away from zero.

\begin{Assumption}
\label{assum:covariate:x:bounded}
The covariates $\{\x_i\}_{i\in[n]}$ satisfy $\sup_{i\in[n]}\|\x_i\|_2 = \mathcal{O}(p^{1/2})$.
\end{Assumption}

Assumption~\ref{assum:covariate:x:bounded} can be satisfied, for example, when $\x$ is standardized and lies in a compact space $[0,1]^p$. Such a boundedness condition is standard in asymptotic analyses of semiparametric and nonparametric regression problems (e.g., \citealt{stone1985additive,huang1999efficient,wang2011estimation,cheng2015joint}).

We first present the convergence rates of the SABRE estimators $\wh{\bb}$ and $\wh{\phi}$ in Theorem~\ref{thm:rate:beta_phi_hat}.  

\setcounter{Theorem}{0}
\renewcommand{\theTheorem}{\arabic{Theorem}} 

\begin{Theorem}[Convergence rates of $\wh{\bb}$ and $\wh{\phi}$]
\label{thm:rate:beta_phi_hat}
Under
Assumptions~\ref{assum:spline:1} to \ref{assum:spline:4}, \ref{assum:p} to \ref{assum:para_space}, and \ref{assum:covariate:x} to \ref{assum:covariate:z}, we have $\|\wh{\bb} - \bb_0\|_2 = \mathcal{O}_p(p^{1/2}n^{-1/2})$. 
Additionally under Assumption~\ref{assum:covariate:x:bounded}, we have $\wh{\phi}-\phi_0 = \mathcal{O}_p\{(p+N)^{1/2}n^{-1/2}\}$.
\end{Theorem}

When $p$ is fixed, $\wh{\bb}$ has the classical $\sqrt{n}$-rate. The SABRE estimator $\wh{\phi}$ attains the same convergence rate as the sMLE $\wt{\phi}$ (see Supplement~E). Under $p \ll n^{1/2}$ and $N \ll n^{1/2}$ (Assumptions~\ref{assum:p} and \ref{assum:spline:3}), Theorem~\ref{thm:rate:beta_phi_hat} implies consistency of both $\wh{\bb}$ and $\wh{\phi}$.

We next study the bias of $\wh{\bb}$ and $\wh{\phi}$. Define the rescaling matrix 
\bq \label{eqn:def_bOmega}
    \bOmega \equiv \begin{bmatrix}
    \I_p & \0_{p\times K} \\
    \0_{K\times p} & N^{-1/2} \I_K
    \end{bmatrix},
\eq
where $\I_p$ denotes a $p\times p$ identity matrix and $\0_{p\times K}$ denotes a $p\times K$ zero matrix. We recall that $K$ and $N$ are of the same order. After rescaling, the parametric component $\wh{\bb}$ and the B-spline coefficient $N^{-1/2}\wh{\balpha}$ converge at the same rate. The bias order of $\bOmega\wh{\bg}$ is given in Proposition~\ref{thm:gamma_hat:bias} below. 

\begin{Proposition}[Bias order of $\bOmega\wh{\bg}$]
\label{thm:gamma_hat:bias}
Under Assumptions~\ref{assum:spline:1} to \ref{assum:spline:4}, \ref{assum:p} to \ref{assum:para_space}, and \ref{assum:covariate:x} to \ref{assum:covariate:x:bounded}, we have $\left\|\bOmega\left\{\mathbb{E}(\wh{\bg})-\bg_0\right\}\right\|_\infty = \mathcal{O}(h^q) + \mathcal{O}\left\{(p+N)^{3/2}n^{-3/2}\right\}$. 
\end{Proposition}

\begin{Theorem}[Bias order of $\wh{\bb}$]
\label{thm:beta_hat:bias}
Under the assumptions of Proposition~\ref{thm:gamma_hat:bias}, we have $\|\mathbb{E}(\wh{\bb}) - \bb_0\|_\infty = \mathcal{O}(h^q) + \mathcal{O}\left\{(p+N)^{3/2}n^{-3/2}\right\}$.
\end{Theorem}

Theorem~\ref{thm:beta_hat:bias} shows that the bias of $\wh{\bb}$ has two components: an $\mathcal{O}(h^q)$ term from approximating-vs-true model discrepancy, which is the price of using a parametric approximating model and implies that a crude approximating model can harm SABRE's accuracy, and an $\mathcal{O}\{(p+N)^{3/2}n^{-3/2}\}$ term from using the sMLE as the initial estimator.

Although no alternative bias-corrected estimators are available for GPLMs, we can compare our bias order to parametric bias correction results with fixed dimension. Ignoring the B-spline component and fixing $p$, the bias of $\wh{\bb}$ reduces to $\mathcal{O}(n^{-3/2})$, which is faster than the $o(n^{-1})$ bias achieved by adjusted-score methods (e.g., \citealt{Firt:93,KoFi:09}). This comparison is informal, since our derivation explicitly accounts for spline approximation and a fully parametric setting might yield sharper results, but it provides a useful benchmark and highlights the competitiveness of SABRE.

To better interpret the bias order of $\wh{\bb}$, we present Corollary~\ref{cor:beta_hat:bias}. 

\begin{Corollary}
\label{cor:beta_hat:bias}
Under the assumptions of Theorem~\ref{thm:beta_hat:bias}, when $N\gg n^{3/(4q)}$ we have $ \|\mathbb{E}(\wh{\bb}) - \bb_0\|_\infty = o(n^{-3/4})$. 
\end{Corollary}

Thus, $\wh{\bb}$ attains a smaller bias order than the standard $o(n^{-1/2})$ bias of existing semiparametric estimators (e.g., the sMLE), whose bias is dominated by the standard error.

\begin{Theorem}[Bias order of $\wh{\phi}$]
\label{thm:phi_hat:bias:main_paper}
Under Assumptions~\ref{assum:spline:1} to \ref{assum:spline:4}, \ref{assum:p} to \ref{assum:para_space}, and \ref{assum:covariate:x} to \ref{assum:covariate:x:bounded}, we have $\mathbb{E}(\wh{\phi}) - \phi_0 = \mathcal{O}\{(p+N)^{1/2}\log(n) h^q\} + \mathcal{O}\{(p+N)^2\log(n) n^{-3/2}\}$.
\end{Theorem}

Theorem~\ref{thm:phi_hat:bias:main_paper} shows that the bias of $\wh{\phi}$ has two components: an $\mathcal{O}\{(p+N)^{1/2}\log(n)\,h^q\}$ term from the approximating-vs-true model discrepancy, and an $\mathcal{O}\{(p+N)^2\log(n)\,n^{-3/2}\}$ term from using the sMLE $\wt{\phi}$ as the initial estimator. In Supplement~E we show that $\wt{\phi}$ itself has bias $\mathcal{O}\{(p+N)^{3/2}n^{-1}\}$, which is larger than both terms above, and thus SABRE substantially reduces the bias relative to sMLE.

While there is limited literature on bias correction for the dispersion parameter, we compare to \citet{saha2005bias} which studies a parametric two-parameter negative binomial distribution and obtains $\mathcal{O}(n^{-1})$ bias for the MLE of the dispersion and $o(n^{-1})$ for their bias-corrected estimator. Ignoring the spline component and fixing $p$, our bias for $\wt{\phi}$ matches their MLE rate, while $\wh{\phi}$ attains the smaller $\mathcal{O}(n^{-3/2})$ bias, highlighting the strength of our approach even in this parametric benchmark.

We now turn to the limiting distribution of $\wh{\bg}$. As implied in \eqref{eqn:decomposition_gamma_hat}, it is closely related to the one of $\wt{\bg}^*$, i.e., the MLE for the B-spline approximating GLM at $(\bg_0,\phi_0)$. Under regularity conditions, $\wt{\bg}^*$ is asymptotically normal with covariance matrix $\bSig_n$.

\begin{Proposition}[Asymptotic normality of $\bOmega\wh{\bg}$]
\label{thm:gamma_hat:asymp_norm}
Under Assumptions~\ref{assum:spline:1} to \ref{assum:spline:4}, \ref{assum:p} to \ref{assum:para_space}, and \ref{assum:covariate:x} to \ref{assum:covariate:z}, for any $\s\in\real^{p+K}$ such that $\|\s\|_2=1$, we have
\bsq
    \sigma_{n,\s,\bg}^{-1} \sqrt{n}\s\trans\bOmega(\wh{\bg}-\bg_0) \overset{D}{\to} \mathcal{N}(0,1) \quad \text{with} \quad \sigma_{n,\s,\bg}^2 \equiv n\s\trans\bOmega \bSig_n \bOmega\s.  
\esq
\end{Proposition}

Therefore, $\wh{\bg}$ and $\wt{\bg}^*$ share the same asymptotic distribution. This result immediately yields the marginal asymptotic distribution of $\wh{\bb}$.

\begin{Theorem}[Asymptotic normality of $\wh{\bb}$]
\label{thm:beta_hat:asymp_norm}
Under the assumptions of Proposition~\ref{thm:gamma_hat:asymp_norm}, for any $\s\in\real^p$ such that $\|\s\|_2=1$ we have
\bsq
    \sigma_{n,\s,\bb}^{-1} \sqrt{n}\s\trans (\wh{\bb}-\bb_0) \overset{D}{\to} \mathcal{N}(0,1) \quad \text{with} \quad \sigma_{n,\s,\bb}^{2}\equiv n\phi_0 \s\trans \left[\wt{\X}\trans \left\{\I_n - \wt{\B}(\wt{\B}\trans\wt{\B})^{-1}\wt{\B}\trans\right\} \wt{\X}\right]^{-1} \s,
\esq
where $\wt{\X}$ is a $n\times p$ matrix with the $i\th$ row to be $b''\{\mu_i^*(\bg_0)\}^{1/2}\x_i\trans$, and $\wt{\B}$ is a $n\times K$ matrix with the $i\th$ row to be $b''\{\mu_i^*(\bg_0)\}^{1/2}\B(z_i)\trans$. 
\end{Theorem}

The SABRE estimator $\wh{\bb}$ shares the same limiting distribution as the sMLE $\wt{\bb}$. Because $\wt{\bb}$ is semiparametric efficient, so is $\wh{\bb}$. Theorem~\ref{thm:beta_hat:asymp_norm} also justifies standard inference based on $\wh{\bb}$, including CIs and hypothesis
tests.

Lastly, although our primary focus is on $\bb$ and $\phi$, the joint analysis of $\wh{\bg}=(\wh{\bb}\trans,\wh{\balpha}\trans)\trans$ also yields asymptotic properties for $\wh{\balpha}$ and the induced estimator $\wh m(\cdot)$, which may be of independent interest. When $m_0(\cdot)$ is the main target, we further propose an \textit{optimal SABRE} estimator that attains the standard convergence rate, minimum MSE, and asymptotic normality for $m_0(\cdot)$ without requiring undersmoothing, while still providing finite-sample bias correction and improved inference. See Supplement~B for details.

\section{Simulation Studies}
\label{sec:simulations}

In this section, we evaluate the finite-sample performance of SABRE. Section~\ref{sec:simulations_beta} studies estimation and inference for $\bb_0$ in partially linear logistic regression with misclassified responses, motivated by medical and epidemiologic settings where binary outcomes (e.g., disease status or exposure indicators) are routinely subject to misclassification. Section~\ref{sec:simulations_phi} examines estimation of $\phi_0$ in partially linear inverse Gaussian and negative binomial regression, two standard models for overdispersed count data in which the dispersion parameter is challenging to estimate, especially when $p/n$ is large and/or dispersion is high.

\subsection{Estimation and Inference of the Parametric Component}
\label{sec:simulations_beta}

We consider a partially linear logistic regression
\bq \label{eqn:logistic_model}
    Y_i^\circ | \x_i,z_i \sim \text{Bernoulli}\left[\text{logit}^{-1}\{\x_i\trans\bb_0 + m_0(z_i)\}\right],
\eq
with $\bb_0 = (-2,-2,-2,-2,4,4,4,4,0.2,-0.2,\ldots,0.2,-0.2)\trans$, $m_0(z) = \sin(5z)$, and $\phi_0$ known to be one. The covariates $\{\x_i\}_{i\in[n]}$ are iid realizations from a $\mathcal{N}\{\0_p, p^{-1}\bSig\}$ with $\Sigma_{kl} = 0.5^{|k-l|}$, and $\{z_i\}_{i\in[n]}$ are iid $U(0,1)$. Instead of $Y_i^\circ$, we observed a misclassified response $Y_i$ with known false positive rate (FPR) $\delta$ and false negative rate (FNR) $\delta/2$, over a grid $\delta \in \{0,0.01,\dots,0.06\}$. This design reflects asymmetric misclassification common in practice, and allows us to use a single parameter $\delta$ to vary both rates. Our results are robust to other configurations. We consider three settings with decreasing $p/n$ ratios: (i) $(p,n)=(60,600)$, (ii) $(90,1200)$, (iii) $(120,2400)$. To approximate $m_0(\cdot)$, we use cubic B-splines with $N$ interior knots at equally spaced quantiles of $\{z_i\}_{i\in[n]}$. We run $10^4$ Monte Carlo replications per setting.

Three estimators for $\bb_0$ are compared: (i) sMLE in \eqref{eqn:def:gamma_phi_tilde} with $N =\text{floor}(n^{1/5})$, (ii) SABRE in \eqref{eqn:def:gamma_phi_hat} with $N =\text{floor}(n^{4/15})$ which satisfies Assumption~\ref{assum:spline:3}, and (iii) sMLE-BR, a bias-reduced sMLE obtained by applying the adjusted-score method of \citet{kosmidis2020mean} (via the \texttt{brglm2} R package) to the B-spline approximating model, with the same B-spline specification as SABRE for comparability. Because this approach requires model-specific analytical derivations, misclassification adjustments are not available.

\begin{figure}[!tb]
    \centering
    \includegraphics[width=0.8\textwidth]{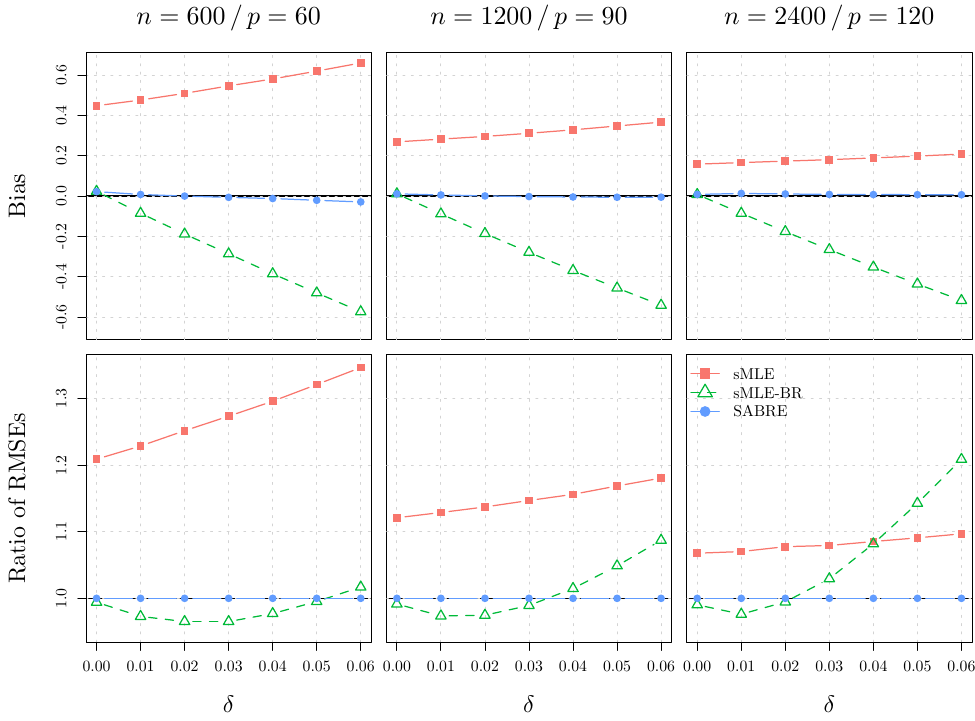}
    \caption{Estimation performance for $\beta_6$ in the partially linear logistic regression simulation in Section~\ref{sec:simulations}. ``Ratio of RMSEs'' refers to the ratio of RMSE of each estimator relative to SABRE.}  
    \label{fig:logistic_misclassification}
\end{figure}

Figure~\ref{fig:logistic_misclassification} summarizes the bias and root MSE (RMSE) for $\beta_6$ with true value 4; other components behave similarly and hence are omitted. When there is no misclassification ($\delta=0$), SABRE and sMLE-BR have negligible bias and lower RMSE than sMLE. Under misclassification ($\delta>0$), sMLE-BR is inconsistent for estimating $\bb$, while sMLE remains consistent but increasingly biased as $\delta$ and/or the $p/n$ ratio grows. SABRE maintains negligible bias across all $\delta$ and $p/n$ regimes, and achieves the smallest or close to the smallest RMSE overall.

\begin{figure}[!tb]
    \centering
    \includegraphics[width=0.8\textwidth]{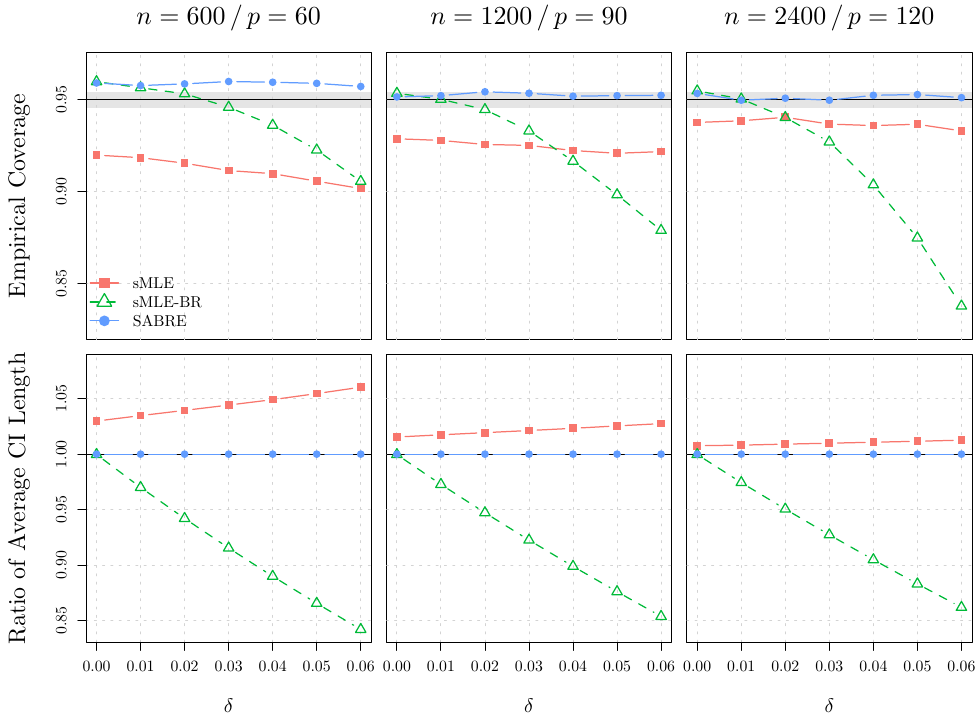}
    \caption{Inference performance for $\beta_6$ in the partially linear logistic regression simulation in Section~\ref{sec:simulations}. The gray zone in the graphs in the first row visualizes the estimated simulation error. ``Ratio of Average CI Length'' refers to the ratio of average CI length of each estimator relative to SABRE.}  
    \label{fig:logistic_misclassification_inference}
\end{figure}

Figure~\ref{fig:logistic_misclassification_inference} reports empirical 95\% CI coverage and average CI length, where the asymptotic variance is estimated by plugging in the point estimates since it admits a closed-form expression. The sMLE has poor coverage, especially at larger $p/n$, primarily due to its substantial bias and resulting inaccurate plug-in variances. sMLE-BR has good coverage when $\delta=0$ but deteriorates quickly under misclassification. In contrast, SABRE delivers coverage close to the nominal level and consistently shorter CIs than competitors, across varying misclassification levels and $p/n$ regimes. 

Estimation and inference results for the nonparametric component $m_0(\cdot)$ are provided in Supplement~C, where SABRE achieves the best performance among all estimators in terms of bias and pointwise CI coverage.

Overall, this study highlights the central role of bias in finite-sample inference: even efficient estimators such as the sMLE can carry enough bias to severely distort coverage, especially at large $p/n$. Parametric bias-correction methods (sMLE-BR) can be force-fit to semiparametric models, but they generally lack theoretical support and do not readily extend to more complex settings such as misclassification. By contrast, SABRE reduces bias, improves RMSE, and yields CIs with more accurate coverage and shorter length across all regimes examined.

\subsection{Estimation of the Dispersion Parameter}
\label{sec:simulations_phi}

We consider a partially linear inverse Gaussian regression 
\bsq
    Y_i|\x_i, z_i \sim \text{IG} \left[ \text{mean} = \{\x_i\trans\bb_0 + m_0(z_i)\}^{-1/2}, \text{shape} = 1/\phi_0 \right],
\esq
with $\bb_0 = (2, -1, -1, -1, 0.1, -0.1, 0.1, -0.1, \ldots)\trans$, $m_0(z) = \sin(5z)+10$, and a grid $\phi_0\in\{0.5,1,1.5,\ldots,4.5\}$. For the covariate vector $\x_i$, its first two components are iid realizations from a $\text{Bernoulli}(0.5)$ and a $\text{Bernoulli}(0.25)$ respectively. The other components of $\x_i$ and $z_i$ are iid $U(0,1)$. Three settings with decreasing $p/n$ ratios are considered: (i) $(p,n)=(50,500)$, (ii) $(75,1000)$, (iii) $(100,2000)$. To approximate $m_0(\cdot)$, we again use cubic B-splines with $N$ interior knots at equally spaced quantiles of $\{z_i\}_{i\in[n]}$. For each setting, we run $10^4$ Monte Carlo replications. 

We compare (i) the sMLE in \eqref{eqn:def:gamma_phi_tilde} with $N = \text{floor}(n^{1/5})$, and (ii) SABRE in \eqref{eqn:def:gamma_phi_hat} with $N = \text{floor}(n^{4/15})$. We also attempted to employ the sMLE-BR, but it failed to converge numerically under the high dispersion levels and large model dimensionalities considered, suggesting that bias correction methods developed for parametric models may not always be directly applicable in these semiparametric settings.

\begin{figure}[!tb]
    \centering
    \includegraphics[width=0.8\textwidth]{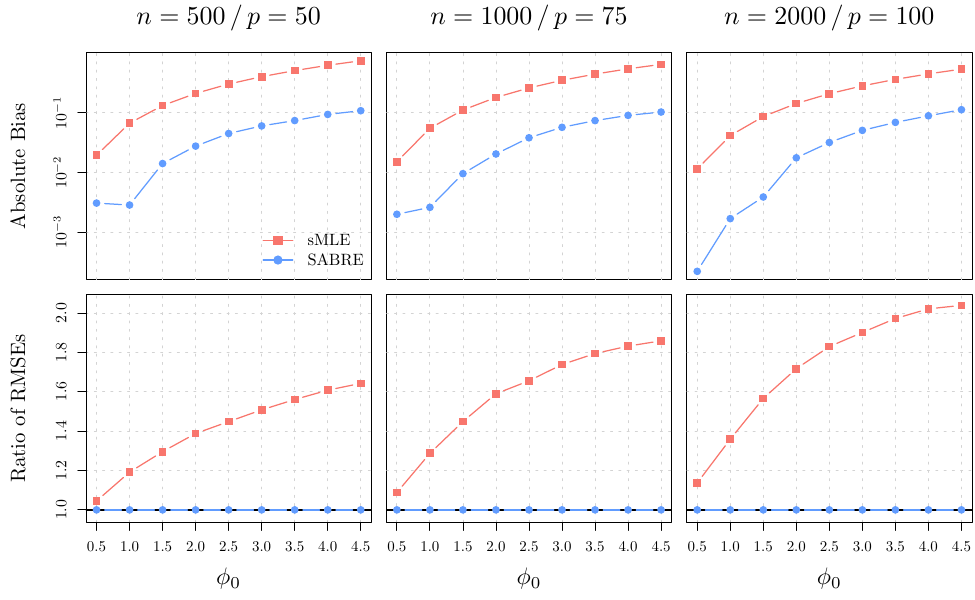}
    \caption{Estimation performance for $\phi_0$ in the partially linear inverse Gaussian simulation in Section~\ref{sec:simulations}. The y-axis of the graphs in the first row is on a log scale. ``Ratio of RMSEs'' refers to the ratio of RMSE of each estimator relative to SABRE.}  
    \label{fig:inv_gauss}
\end{figure}

Both sMLE and SABRE show negligible bias and similar RMSE for estimating $\bb_0$, so we focus on  the results for $\phi_0$ as summarized in Figure~\ref{fig:inv_gauss}. The top row shows that sMLE has substantial bias, whereas SABRE reduces the bias down to negligible. Bias increases for both methods as dispersion level and/or the $p/n$ ratio grow, but SABRE always has smaller bias. The bottom row reports that the RMSE of sMLE relative to SABRE exceeds 1 in all settings, indicating uniformly smaller RMSE for SABRE. Therefore, SABRE is particularly advantageous at reducing the estimation bias for $\phi_0$, especially in challenging settings with large
dispersion and model dimensionality.

In Supplement~C, we report simulations for a partially linear negative binomial regression, comparing sMLE, sMLE-BR, and SABRE. Across all settings, the estimation of $\bb_0$ shows negligible bias, whereas $\phi_0$ is much harder to estimate. When estimating $\phi_0$, sMLE exhibits substantial bias, especially under high dispersion and large $p/n$; both sMLE-BR and SABRE reduce this bias, with SABRE achieving the smallest bias and the lowest RMSE for $\phi_0$ among all methods.

\section{Alzheimer's Disease Genetics Association Analysis}
\label{sec:case_study}

Alzheimer's disease (AD) is the most common form of dementia and contributes to $60$--$70\%$ of cases \citep{who_dementia}. AD is highly heritable, and identifying the genetic variants that shape its risk both illuminates disease biology and supports genetic risk prediction. Large genetic studies of AD increasingly rely on EHR-linked biobanks, where gold-standard disease labels are unavailable at scale. Instead, AD status is often represented by diagnostic codes or EHR-based phenotyping algorithms, both of which are subject to misclassification. We use data from the Mass General Brigham (MGB) Biobank to examine whether SABRE can reduce bias in such AD genetic association studies in the presence of error-prone outcomes.

The study includes $38{,}471$ participants aged at least $55$ at their last EHR encounter. We first used the presence of any dementia-related diagnosis code as a screening filter, identifying $2{,}058$ filter-positive and $36{,}413$ filter-negative participants. Under the study design, filter-negative participants were treated as true AD-negative controls because they had no recorded dementia-related code. Within the filter-positive group, we defined a surrogate AD label based on a knowledge-driven online multimodal automated phenotyping (KOMAP) algorithm score at a cutoff attaining 90\% specificity \citep{xiong2023knowledge,venkatesh2026leveraging}, yielding $864$ surrogate-positive and $1{,}194$ surrogate-negative participants. We retained all filter-positive participants and randomly sampled $4{,}320$ filter-negative controls (five times the number of surrogate-positive participants), resulting in an analytic cohort of $n = 6{,}378$. Based on chart review of $100$ randomly selected filter-positive participants, this surrogate has an estimated FPR of $0.019$ and FNR of $0.181$ in the analytic cohort. Details of the cohort construction and of the estimation of these two rates are provided in Supplement~D.

Our main goal is to determine to what extent the well-established genetic architecture of AD can be recovered using this EHR-linked biobank data with imperfect AD outcome annotation. In particular, the APOE-$\varepsilon 4$ allele strongly increases AD risk, whereas APOE-$\varepsilon 2$ is protective \citep{corder1993gene,farrer1997effects}. Prior genome-wide association studies (GWAS) also identified a number of other variants potentially related to AD \citep{jansen2019genome}. Because dyslipidemia, particularly elevated LDL cholesterol, may contribute to AD through vascular and neurodegenerative pathways, we include LDL-associated variants alongside established AD variants \citep{reitz2013dyslipidemia,van2021genetic}. After quality control (minor-allele frequency $\ge 5\%$, per-SNP missingness $< 5\%$) and linkage disequilibrium (LD) pruning to a near-independent set (pairwise genotype squared correlation $r^2< 0.10$; see e.g., \citealt{anderson2010data}), we order the variants by GWAS Catalog trait tier, taking those mapped within $100$~kb of a genome-wide significant association with an AD or dementia trait first, then with a lipid trait, then with a cardiovascular trait, and breaking ties within a tier by genotype variance; we then retain the top $169$ SNPs. Together with the two APOE allele dosages, sex, and race, this gives a parametric dimension of $p=173$ and an events-per-variable ratio of approximately five, a setting in which finite-sample bias can be substantial. Age is modeled nonparametrically using cubic B-splines with the number of knots selected by cross-validation. See Supplement~D for more details about the cohort composition.

For subject $i$ with covariates $\x_i$ (including the two APOE dosages, SNPs, sex, and race) along with age$_i$, we model the AD risk via a partially linear logistic regression,
\bsq
    \pr(Y_i^\circ=1\mid \x_i,\text{age}_i) = \text{logit}^{-1}\left\{\x_i\trans\bb + m(\text{age}_i)\right\},
\esq
with $Y_i^\circ$ the latent true AD status, $\bb$ the parameter of interest, and $m(\cdot)$ a smooth nuisance function. Given the FPR and FNR of the surrogate, the observed label $Y_i$ satisfies
\bq \label{eqn:ad_misclass}
    \pr(Y_i=1\mid \x_i,\text{age}_i) = \mathrm{FPR} + (1-\mathrm{FPR}-\mathrm{FNR})\,\pr(Y_i^\circ=1\mid \x_i,\text{age}_i).
\eq
We fit this model using both sMLE and SABRE, with plug-in estimates of their asymptotic variances used to construct CIs. 

Table~\ref{tab:ad_real} reports the estimated APOE odds ratios (ORs) with $95\%$ CIs, alongside the per-allele ORs from the meta-analysis of \cite{farrer1997effects}, which we take as an external benchmark. Both estimators recover the protective $\varepsilon 2$ association, with ORs of $0.63$ for sMLE and $0.66$ for SABRE, both consistent with the benchmark. The two estimators differ materially for $\varepsilon 4$: SABRE gives an OR of $3.58$, inside the benchmark CI, whereas sMLE gives $4.05$, above it. At an events-per-variable ratio of about five and with a misclassified outcome, this is the direction of distortion that our theory predicts for sMLE, and the emulation study below confirms it.

The remaining covariates show little evidence of association. Sex (OR $0.95$) and race (OR $0.89$ for non-white) are non-significant under both estimators. All $169$ SNP ORs lie between $0.78$ and $1.22$ for SABRE and between $0.76$ and $1.25$ for sMLE, and only $11$ SABRE CIs and $13$ sMLE CIs exclude one, close to the number expected by chance. Full results are reported in Supplement~D.

\begin{table}[!tb]
\centering
\caption{Estimated APOE odds ratios (ORs) with $95\%$ CIs in the MGB Biobank AD analysis, under the KOMAP surrogate at $90\%$ specificity. The last column reports the external benchmark of \cite{farrer1997effects}.}
\label{tab:ad_real}
\begin{tabular}{lccc}
\toprule
 & sMLE & SABRE & Benchmark \\
Allele & OR ($95\%$ CI) & OR ($95\%$ CI) & OR ($95\%$ CI) \\
\midrule
APOE-$\varepsilon 4$ & $4.05\ (3.35,\ 4.89)$ & $3.58\ (2.99,\ 4.28)$ & $3.2\ (2.8,\ 3.8)$ \\
APOE-$\varepsilon 2$ & $0.63\ (0.44,\ 0.89)$ & $0.66\ (0.47,\ 0.91)$ & $0.6\ (0.5,\ 0.8)$ \\
\bottomrule
\end{tabular}
\end{table}

\begin{figure}[!tp]
    \centering
    \includegraphics[width=\textwidth]{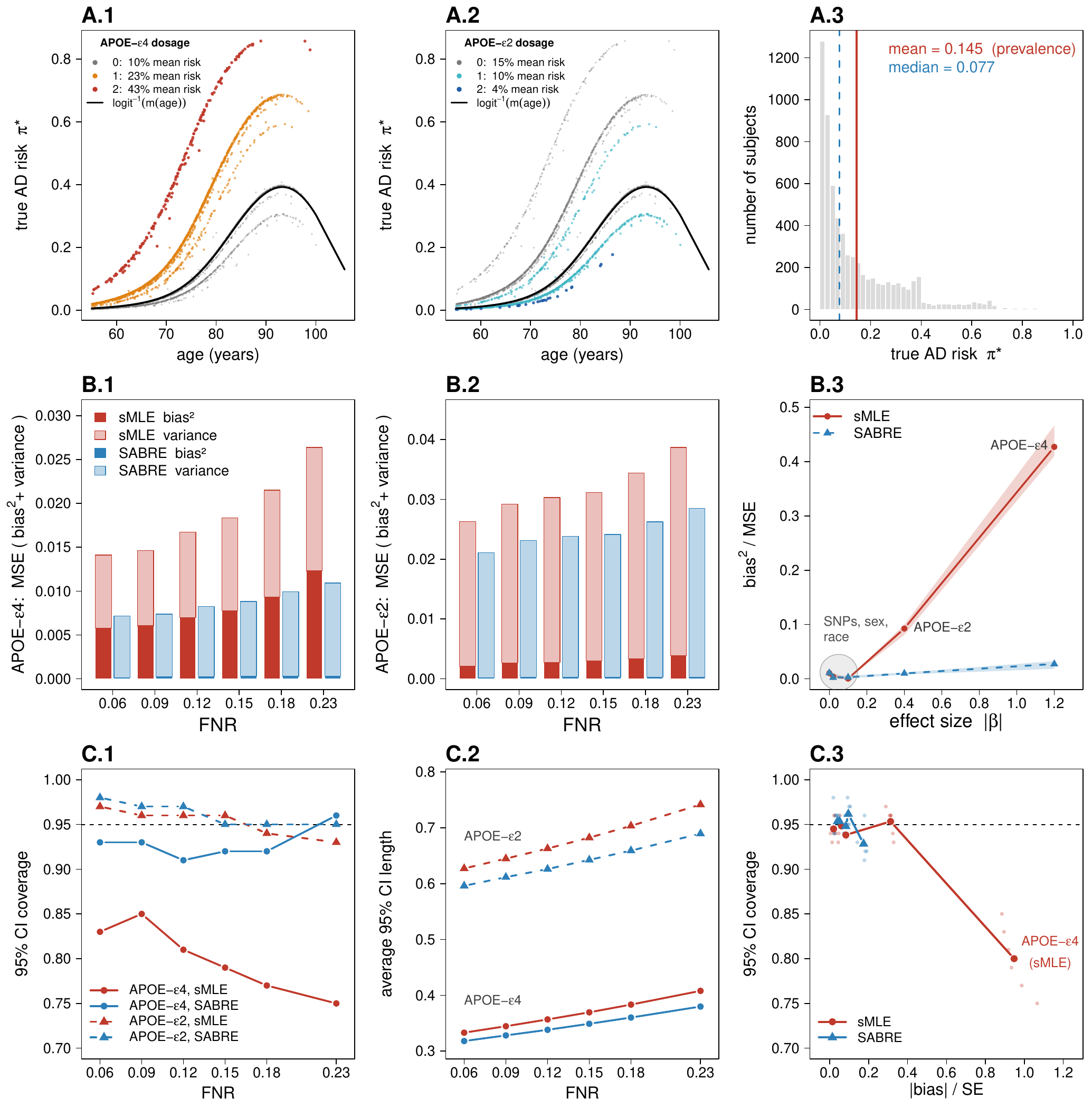}
    \caption{Known-truth emulation validating the real-data AD genetics association analysis in Section~\ref{sec:case_study}. \textbf{Row A}, data-generating mechanism: true AD risk $\pi^*$ against age by $\varepsilon 4$ (A.1) and $\varepsilon 2$ (A.2) dosage, with the distribution of $\pi^*$ (A.3). \textbf{Row B}, estimation: MSE as squared bias (solid) plus variance (faded) against the FNR, for $\varepsilon 4$ (B.1) and $\varepsilon 2$ (B.2), and the squared-bias share against effect size (B.3). \textbf{Row C}, inference: $95\%$ CI coverage (C.1) and average CI length (C.2) against the FNR, and coverage against standardized bias (C.3).}
    \label{fig:ad_emulation}
\end{figure}

\paragraph{Known-truth Emulation Study} Because the true genetic effects are unknown in the real data, we conduct an emulation study calibrated to the real-data analysis. We set the true APOE ORs to $3.32$ for $\varepsilon 4$ and $0.67$ for $\varepsilon 2$, assign small effects to sex and race, and set all 169 additional SNP effects to zero. The age function and intercept are calibrated to reproduce the observed age pattern and event rate. See Supplement~D for more details about the study design. The resulting latent AD prevalence is 0.145, with mean risks increasing from 10\% to 23\% and 43\% across zero, one, and two $\varepsilon 4$ alleles, and decreasing from 15\% to 10\% and 4\% across the corresponding $\varepsilon 2$ dosages; see Figure~\ref{fig:ad_emulation}(A).

We generate 100 datasets with $\mathrm{FPR}=0.019$ and $\mathrm{FNR}$ ranging from 0.06 to 0.23. Figure~\ref{fig:ad_emulation}(B) shows that the MSE of both estimators increases as misclassification worsens, but SABRE has lower MSE at every $\mathrm{FNR}$. The difference is driven primarily by bias: the squared bias of sMLE increases sharply with the $\mathrm{FNR}$, whereas SABRE's bias remains negligible. The contribution of bias to MSE also increases with effect size, explaining why the distortion is most pronounced for the large $\varepsilon 4$ effect, more moderate for $\varepsilon 2$, and minimal for the small or null effects.

Figure~\ref{fig:ad_emulation}(C) shows the corresponding inferential performance. The sMLE coverage for $\varepsilon 4$ declines to 0.75 under severe misclassification, whereas SABRE remains close to the nominal 95\% level and produces shorter CIs. Coverage for the smaller $\varepsilon 2$ effect is less affected. Across parameters and surrogate qualities, coverage is closely determined by the standardized bias $|\mathrm{bias}|/\mathrm{SE}$: sMLE coverage deteriorates as this ratio increases, whereas SABRE keeps it near zero.

In summary, the real-data and emulation analyses show that finite-sample bias can substantially inflate major genetic associations even when the estimator is asymptotically efficient. SABRE corrects this inflation, recovers the established risk-increasing and protective APOE effects, and provides more accurate inference in a realistic large-scale genetic association analysis with outcome misclassification.

\section{Conclusions}
\label{sec:conclusion}

Bias plays a central role in finite-sample inference, yet the semiparametric literature on models of the form $f\{Y|\x\trans\bb+m(z),\phi\}$ has largely focused on semiparametric efficiency for $\bb$ in specific model subclasses, treating $\phi$ and $m(\cdot)$ as nuisances and overlooking finite-sample bias. To address this gap, we have proposed SABRE, a simulation-based bias correction framework for this broad semiparametric model class that also allows a diverging parameter dimension. SABRE leverages a parametric approximating model built from a B-spline approximation of $m(\cdot)$: starting from the B-spline sMLE, SABRE is defined by matching this initial estimator to its simulation-based expectation under the same approximating model.

Within the GPLM subclass, we developed a comprehensive asymptotic theory, including convergence rates and smaller bias orders for $\bb$ and $\phi$, and asymptotic normality for $\bb$ with semiparametric efficiency. Our analysis also yielded asymptotic properties for the induced estimator of $m(\cdot)$. We further provide a roadmap indicating how the theory could, in principle, be extended to the broader model class, but do not pursue the full generality here, as the derivations are highly distribution-specific.

Simulation studies demonstrated substantial finite-sample gains of SABRE over existing approaches. Across partially linear logistic, inverse Gaussian, and negative binomial regression models, SABRE markedly reduced bias in $\bb$ and $\phi$, with the most pronounced gains when $p/n$ is large, misclassification is severe, or dispersion is high. These bias reductions yielded CIs for $\bb$ with more accurate empirical coverage and shorter length. We also applied SABRE to an Alzheimer's disease genetics dataset from an EHR-linked biobank with an accompanying emulation study, where it delivered shorter CIs, improved coverage, and recovered the disease's well-established genetic risk factors while correcting the inflated effect sizes that standard methods produce, thereby demonstrating its practical effectiveness and value for scientific interpretation.

\section*{Data Availability Statement} 
Individual-level patient data from the Mass General Brigham Biobank cannot be shared due to privacy protections and institutional constraints; access may be requested by qualified investigators through the Biobank's governance process. GWAS Catalog annotations are publicly available at \url{https://www.ebi.ac.uk/gwas/}.

\section*{Disclosure Statement} 
There are no relevant financial or non-financial competing interests to report.

{\renewcommand{\baselinestretch}{1.7}\small
\normalem 
\bibliographystyle{apalike}
\bibliography{refs}
}

\end{document}